# Tactile Materials in Practice: Understanding the Experiences of Teachers of the Visually Impaired

MAHIKA PHUTANE

Cornell University, mahika@cs.cornell.edu

JULIE WRIGHT

Portland State University, wrig6@pdx.edu

BRENDA VERONICA CASTRO

Hunter College, bvcastro42@gmail.com

LEI SHI

Cornell Tech, Cornell University, ls776@cornell.edu

SIMONE R. STERN

Swarthmore, sstern2@swarthmore.edu

HOLLY M. LAWSON

Portland State University, hlawson@pdx.edu

SHIRI AZENKOT

Jacobs Technion-Cornell Institute, Cornell Tech, shiri.azenkot@cornell.edu

Teachers of the visually impaired (TVIs) regularly present tactile materials (tactile graphics, 3D models, and real objects) to students with vision impairments. Researchers have been increasingly interested in designing tools to support the use of tactile materials, but we still lack an in-depth understanding of how tactile materials are created and used in practice today. To address this gap, we conducted interviews with 21 TVIs and a 3-week diary study with eight of them. We found that tactile materials were regularly used for academic as well as non-academic concepts like tactile literacy, motor ability, and spatial awareness. Real objects and 3D models served as "stepping stones" to tactile graphics and our participants preferred to teach with 3D models, despite finding them difficult to create, obtain, and modify. Use of certain materials also carried social implications; participants selected materials that fostered student independence and allow classroom inclusion. We contribute design considerations, encouraging future work on tactile materials to enable student and TVI co-creation, facilitate rapid prototyping, and promote movement and spatial awareness. To support future research in this area, our paper provides a fundamental understanding of current practices. We bridge these practices to established pedagogical approaches and highlight opportunities for growth regarding this important genre of educational materials.

• **Human-centered Computing** ~ Accessibility • **Applied Computing** ~ Education

## 1 INTRODUCTION

Tactile materials allow enriching educational experiences for students with vision impairments and are used regularly by teachers of the visually impaired (TVIs) to convey concepts. However, tactile material preparation can be a complex, time-consuming, and labor-intensive process [48,69,70]. Teachers have stressed the challenges involved in designing meaningful tactile graphics that are simple, straightforward, and represent information in a sequential way, similar to how one processes visual information [48]. Tactile graphics can be overwhelming with tactile stimuli and information, and students have reported problems regarding confusing diagrams and inadequate braille labels [2,45,71].

Technologies have addressed these challenges by streamlining the creation of tactile materials, such as automating the "translation" of visual images to tactile graphics [5,32,42], converting table-based coordinate spaces to tactile formats [7,8], and developing systems to 3D print high-resolution models [10,18,38]. Researchers have also mitigated concerns surrounding braille labels by incorporating audio with the tactile material. Baker et al. [4] embedded QR codes into tactile graphics to provide audio descriptions, and in our previous work [50,52], we developed a toolkit that allows people to add audio annotations to 3D models. Researchers continue to design tactile materials and systems to further education in specific subject areas, including mathematics [21,26,33], computer science [16,25], and the natural sciences [24].

While these innovations ease some of the challenges surrounding tactile materials and show promise in lab studies, the use and practice of tactile materials in the classroom remains unaddressed. In order to create tools that teachers would prefer to use, and that better suit the needs of students with vision impairments, we must study the current practice of tactile materials. Gaining a deeper understanding of how teachers use tactile materials with their students, and how they make decisions about which kinds of tactile materials to present, would assist future researchers in creating tools that appropriately address the needs and challenges of teachers and their students in real-use contexts.

Teachers present tactile materials as either tactile graphics, 3D models, or real objects, but most research has focused on tactile graphics. *Tactile graphics* refer to raised surfaces and line drawings that convey non-text visual information such as maps, graphs, and diagrams. *3D models* refer to any objects that are representative of real-world or abstract concepts and include items such as toy cars, plastic animal replications, and abacuses. *Real objects,* or *realia,* refer to unamended surfaces, textures, and items found in the environment such as cutlery, tree leaves, and fabric [66]. Teachers make constant decisions about which types of materials to use and how to present them to their students. For example, if their student is reading a picture book about a koala, they may present tactile graphics that convert the animal images to raised line drawings, or they may purchase a 3D model of a koala in its habitat, or they may even bring faux fur and leaf textures to help the student engage with the story. Previous studies with TVIs focused on the creation and use of tactile graphics [2,48,69,71], whereas instructional methods with 3D models and real objects remain unexplored. Despite the growth of technologies to support tactile material instruction, there has been no investigation to understand the current practice of tactile materials.

To address these gaps, we pose the following research questions:
- What purpose do tactile materials serve?
- How do TVIs create and acquire tactile materials for their students?
- What tools do TVIs use to support their process?
- Why do TVIs use different types of tactile materials (tactile graphics, 3D models, real objects)?

To answer these questions, we conducted (1) semi-structured interviews with 21 TVIs and (2) a diary study with eight of them. With interviews, we gathered a contextual understanding of tactile material creation and acquisition, as described by TVIs themselves. Specifically, we identified the factors that influenced their use of tactile materials and gained insight into the challenges they faced. With the diary study, we received detailed information about the tactile materials that TVIs presented to their students. The images and responses attached to each diary entry revealed how the TVIs selected, prepared, and adapted tactile materials for their students.

We found that teachers used tactile materials for reasons beyond conveying academic content, such as building their students' tactile literacy, motor abilities, and spatial sense. Teachers opted for materials that students would find engaging and simulating; incidentally, such materials often involved student movement. Teachers preferred real objects and 3D models as teaching aids, especially for foundational concepts, but found them obtrusive and difficult to create or obtain. Hence, they often settled for non-durable handmade materials or resorted to tactile graphics. Tactile materials also carried social implications, and teachers gravitated towards materials that fostered student self-determination and classroom inclusion. Many teachers did not rely on technology to create tactile materials; they found machines cumbersome, expensive, or inadequate for representing concepts they wanted to convey.

Given the growing interest in assistive technologies for tactile material creation, our research provides a fundamental understanding of current practices, highlighting considerations and opportunities for continued research to support teachers and students with visual impairments.

## 2 RELATED WORK

We describe prior work on understanding tactile material use and examine traditional and newer methods of tactile material creation. We then discuss literature on educational practices for students with vision impairments. Our research is the first to investigate the full range of tactile materials (tactile graphics, 3D models, real objects), address concerns around technology adoption by TVIs, and outline emerging gaps between tactile material research and tactile materials in practice.

### 2.1 Understanding the Use of Tactile Materials

Some researchers have investigated how tactile graphics—a subset of tactile materials—are used by students and teachers. Through focus groups and surveys, researchers discovered that students with visual impairments found complex tactile graphics to decrease their interest and enthusiasm in learning concepts. Misplaced braille labels, and overwhelming tactile stimuli were common complaints by students [2,45,71]. Sheppard and Aldrich conducted surveys with TVIs to gather their perspectives on the design and use of tactile graphics [48]. Teachers emphasized both, the importance of tactile graphics as educational tools and the difficulties with them: production was labor intensive, and cluttered graphics were hard to make meaningful to learners.

Accessibility researchers have considered 3D printed models as tactile materials for various STEM subjects [3,24,25,28,62], yet little investigation has been done to study the effectiveness of 3D models in the classroom. Wright et al. surveyed existing literature on tactile maps and 3D models, and suggested that training students with vision impairments to better grasp tactile materials could enhance their symbolic understanding and spatial sense [67]. Within this context of tactile maps, Holloway et al. [20] were the first to compare the use of tactile graphics and 3D models. Their findings dictate that 3D models were preferred, and especially offered an advantage for orientation and mobility training.

Prior studies have examined TVIs in the context of assistive technology adoption. Recent tactile material production methods (see Section 2.2), especially those involving newer forms of software and hardware, require TVIs to be technically proficient. Prior work suggests that only 40% of TVIs integrate newer forms of technology into their instructional methods [1], and recent initiatives confirm that assistive technology (AT) integration is an ongoing issue for new tactile material adoption [65]. Siu et al developed a measure to assess the technical proficiency of TVIs with assistive technology [55], emphasizing that a social community of practice [40] is necessary in adoption of newer AT.

In addition to investigating the full range of tactile materials (3D models, tactile graphics, real objects), our study provides guidelines and recommendations for future work that acknowledges the TVIs' resistance to newer forms of technology. Our considerations support researchers in designing technologies that address TVI needs and are likely to be adopted in the context of a classroom.

## 2.2  Tools and Methods to Create Tactile Materials

Rowell and Ungar [46] sought to understand the practices of tactile graphic production and surveyed researchers, educators, and tactile map producers from various different countries. They found no significant patterns and similarities in tactile graphic production across different geographic locations, concluding that creators worked independently and did not share knowledge in a formal way. To this end, our study aims to bridge the gap between tactile material production in *research and innovation*, and tactile material production *in practice*.

The most common traditional tactile graphic production methods involve the use of swell paper and braille embossing [46]. Swell paper contains microcapsules of alcohol that rise when exposed to heat; drawing on swell paper with black ink and exposing these drawings to heat in a Pictures-in-a-Flash (PIAF) Machine causes these markings to swell and creates raised lines drawings. Braille embossers produce raised dots; both these methods can be connected to software that allow TVIs to create and modify their tactile graphics.

Prior research has expedited this tactile graphic production process for specific contexts. For example, Braier et al. [7] and Brown and Hurst [8] created automation tools to create tactile representations of table-based data and coordinate spaces. Way and Barner [60,61] created a system to convert images from visual to tactile form through edge detection, image processing, and the swell paper method. Researchers continued to develop ways to automate the translation of visual images to tactile graphics using more advanced techniques [5,32,42], such as Jayant et al. [22] who implemented machine learning and computational geometry methods to translate textbook images, achieving an average translation time of less than ten minutes. These innovations reduce the overload of creating tactile graphics, however it is uncertain whether TVIs employ these production techniques in the classroom.

Researchers sought to improve comprehension of tactile maps by producing 3D legends [18], or by incorporating multimodal output. Minhat et al. [39] created talking tactile maps that inform the user of their surroundings provide specific directions for reaching their destination. We designed an interactive and iterative tactile map [36,53] that helps students practice orientation and mobility skills. Celani and Milan [10] used laser cutting to successfully, albeit expensively, produce robust, high-resolution tactile models—2.5D maps—that would be less abstract than tactile maps; users found them easier to understand than tactile maps because of additional references, such as an elevator or stairs. Additionally, researchers created web and mobile tools to support users in customizing and 3D printing their tactile map models [17,57].

The popularity of 3D printing technology also expanded the practice of rapid prototyping and production of 3D models. Giraud et al. [16] and Stangl et al. [56] investigated how amateur designers fared in creating their own "Do-It-Yourself" 3D models and found that they needed assistance and would benefit from an online creativity support tool. Studies have shown that creating interactive tactile maps and other 3D models through rapid prototyping tools increased the accessibility of various subjects—such as graphic design theory [36], computer science [16,25], and astronomy [6]. In our prior work [49–52], we developed software that allows users to annotate such 3D models with audio by complementing specific elements of the model with text annotations.

Despite these innovations—automation in tactile graphic production and rapid prototyping in 3D model production—we lack awareness of current production practices of tactile materials by TVIs. Our diary study provides a foundation for current tactile material creation and use patterns by TVIs. Given the concerns regarding tactile material adoption [1,65], we provide design recommendations for future researchers that address TVI needs and ensure that further innovations in this space can be broadly adopted in an educational setting.

### 2.3 Educational Practices for Students with Vision Impairments

To provide a holistic reflection on tactile material use, we study prior literature on educational practices and identify elements that pertain to TVIs and students with vision impairments.

Students with vision impairments follow an additional curriculum—the Expanded Core Curriculum (ECC)—focussing on the development of specific skills that compensate for vision loss. These skills are foundational to all other learning; examples include sensory efficiency (using all available senses efficiently to access information), and self-determination (advocating for one's needs) [13]. Studies have shown that low self-confidence, reduced personal agency, and peer exclusion are concerning issues that detriment the academic success of students with vision impairments [12,19,23]. The ECC is not only a substantial part of TVI instruction, it is critical for the development and success of students beyond the classroom [47,64]. Prior work in the tactile material research community has focussed on STEM subjects [3,24–26,28,33,62], but we lack an understanding of how teachers develop tactile materials to convey concepts from the ECC—a gap this paper aims to address.

Lieberman et al. [34] studied how ECC concepts can be taught through physical education, which leads to the notion of *movement* in educational practice. Existing literature on Embodied Learning (EL) has shown strong correlations between physical movement and improved academic performance [14,30,41], particularly regarding improvements in cognitive, motor, and academic abilities for students with special needs [29,31]. The educational approach of EL has confirmed that manipulatives, gestures, and physical movements play a large role in communicating math concepts [58], and can be especially important for students with no vision [26,33]. For instance, Tran et al. [58] explain that counting the number of dots in a tactile pattern can be a challenging task for children with visual impairments. They need to "make sure that they found all dots and did not count any dot twice, [which] is difficult because touch does not provide a preliminary overview of the pattern the way vision does." Within the scope of tactile materials, Kim and Yeh [27] designed movable tactile graphics for students with vision impairments, however, the intersections of EL and tactile materials remain largely unexplored.

Involvement of participants in the design process—Participatory Design (PD)— and involvement of students in the teaching process—Students as Partners (SaP)—is a parallel we find interesting to explore. Researchers have involved users with vision impairments in the design process of AT [59,63,68]; Pires et al. [43] conducted PD sessions with students and TVIs in the context of math education. In pedagogical literature, Cook-Sather et. al [11] developed the SaP approach, "through which all participants have the opportunity to contribute equally, although not necessarily in the same ways, to curricular or pedagogical conceptualization, decision-making, implementation, investigation, or analysis." This instructional practice is mostly observed in higher education, and studies have shown that a shared sense of responsibility, respect, and trust within students have led to a transformational effect on their participation and confidence [37].

Mostly, these educational practices remain separate from the tactile materials research space, although their intersections are necessary and meaningful to explore. Our study offers deeper insights on how TVIs design tactile materials and incorporate EL and SaP approaches with their students, while acutely being aware of the interpersonal dynamics at play.

## 3 INTERVIEW STUDY: UNDERSTANDING EXPERIENCES OF TVIS WITH TACTILE MATERIALS

### 3.1 Method

We conducted semi-structured interviews with TVIs and braillists to understand their experiences with tactile materials. Our interview study left many unanswered questions (see Section 3.3) and propelled a need for a secondary study with a longitudinal component. Thus, we conducted this study in two parts— an initial interview study with 13 participants, and then an interview *and* a diary study with 8 additional participants (see Section 4: Diary Study). Interview protocols across both parts were kept consistent, so this section will discuss interview findings from all 21 participants.

***Participants***. A total of 21 participants (see Table 1) partook in the interview study, comprising of 20 females and one male. For participants to be eligible for this study, they needed to be over 18 years of age, own a smartphone or tablet, be employed as either a Teacher of the Visually Impaired (TVI) or a braillist in the United States (US), and have students who regularly use tactile materials. Braillists are professionals whose job is specifically to "transcribe" educational materials into alternative formats for students with visual impairments. Not all school districts have braillists, and in most cases, those responsibilities lie with a TVI. We thus included both TVIs and braillists in our study since both create and use tactile materials with students. We restricted our study to US-based participants because educational systems in other countries differ, along with the role and responsibilities of a special education teacher.

Participant ages ranged from 31 to 67 years with a mean of 47.8 years (SD = 12.1). Thirteen of those participants identified themselves as sighted, four as blind, one as low vision, and three chose not to disclose. Their professional experience as a TVI or braillist varied, ranging from one year of experience to more than 30 years. Participation was voluntary, and participants were recruited via email and snowball sampling.

Table 1: Demographic information for participants.

| *Pseudonym* | *Gender/Age* | *Vision* | *Position* | *Experience (in years)* | *No. of Students* | *Characteristics of Current Students* |
|---|---|---|---|---|---|---|
| *Elizabeth* | F/47 | Sighted | Teacher of Visually Impaired | 18 | 21 | Low Vision and Blind, Multiple students had both intellectual and motor disabilities, K-12 |
| *Rebecca* | F/40 | Low Vision | Retired Teacher of Visually Impaired and Independent Contractor | 16 | 2 | All Blind, Rarely a student without multiple disabilities, K-12 |
| *Tiffany* | F/64 | Sighted | Teacher of Visually Impaired and Special Education Teacher | 30+ | 12-21 | All Low Vision, 50% in wheelchairs, Multiple with intellectual, emotional, or hearing disabilities, K-12 |

| Pseudonym | Gender/Age | Vision | Position | Experience (in years) | No. of Students | Characteristics of Current Students |
|---|---|---|---|---|---|---|
| Ashley | F/36 | Blind | Teacher of Visually Impaired | 4 | 8 | Majority Blind, K-12 |
| David | M/34 | Sighted | Teacher of Visually Impaired and Orientation and Mobility (O&M) Instructor | 6 | 13 | Majority Blind, Multiple students with motor and intellectual disabilities, K-12 |
| Lauren | F/67 | Sighted | Teacher of Visually Impaired | 16 | 15 | Low Vision and Blind, No other diagnosed disabilities, K-12 |
| Lisa | F/55 | Sighted | Teacher of Visually Impaired and O&M Instructor | 29 | 4 | Majority Low Vision, 3 with intellectual disabilities, K-12 |
| April | F/59 | Sighted | Teacher of Visually Impaired and O&M Instructor | 38 | 13 | Majority Low Vision, 1 with both an intellectual and motor disability, K-12 |
| Donna | F/58 | Blind | Teacher of Visually Impaired and Retired Rehabilitation Teacher | 17 | 5 | Majority Low Vision, 1 with an intellectual impairment, K-12 |
| Christine | F/52 | Sighted | Teacher of Visually Impaired | 12 | 13 | Majority Low Vision, Multiple with intellectual and motor disabilities, All Ages |
| Jill | F/- | - | Teacher of Visually Impaired | - | 10 | Majority Low Vision, Majority multiple disabilities, K-12 |
| Carolyn | F/- | - | Teacher of Visually Impaired | - | 9 | Low Vision and Blind, 18+ |
| Maria | F/- | - | Teacher of Visually Impaired | - | 15 | Majority Low Vision, 8 have multiple disabilities, K-12 |
| Tahani* | F/51 | Sighted | Braillist | 5.5 | 4 | Low Vision, Multiple disabilities, K-12 |
| Eleanor* | F/32 | Sighted | Teacher of Visually Impaired | 1 | 14 | Low Vision and Blind, Multiple disabilities, K-12 |
| Blossom* | F/37 | Blind | Teacher of Visually Impaired | 5 | 18 | Low Vision and Blind, Multiple disabilities, K-12 |
| Chidi* | F/31 | Blind | Teacher of Visually Impaired | 4 | 23 | Low Vision and Blind, Multiple disabilities, K-12 |
| Alexis* | F/50 | Sighted | Teacher of Visually Impaired, O&M Specialist | 24 | 10 | Low Vision and Blind, Multiple disabilities, K-12 |
| Olivia* | F/58 | Sighted | Teacher of Visually Impaired, O&M Specialist, Vision Rehab Teacher | 16 | 25 | Low Vision and Blind, Multiple disabilities, K-12 |
| Connie* | F/57 | Sighted | Teacher of Visually Impaired | 7 | 8 | Low Vision, Multiple disabilities, K-12 |
| Kamilah* | F/35 | Sighted | Teacher of Visually Impaired, O&M Specialist | 5 | 18 | Low Vision and Blind, Multiple disabilities, K-12 |

*Participant also took part in the diary study. Due to a lapse in communication, we have limited information for Jill, Carolyn, and Maria.

*Procedure*. The interview was conducted via phone or video conference software, recorded, and lasted approximately 45-60 minutes. We asked participants about their vision, demographics, and TVI experience—specifically regarding caseload and role as a TVI or braillist. We asked participants to reflect on what materials they created, adapted, or purchased for their students. A series of questions was used to (1) ascertain when and why the participant would use these materials, (2) when they would use both tactile graphics and 3D models together, (3) how they acquired these materials, and (4) the advantages and disadvantages of tactile graphics and 3D models. Participants provided verbal descriptions and references to online catalogues of tactile materials to detail the materials they used.

*Analysis*. We audio recorded and transcribed all 21 interviews, while assigning pseudonyms to all participants to keep their identities anonymous. The data was analyzed in two rounds by two separate pairs of researchers. The first pair analyzed the first 13 interviews using open coding, formulated a code book, and wrote an initial draft of the findings. The second pair analyzed the initial 13, and the additional 8 interviews; to do so, they recoded all 21 interviews to gain a cohesive understanding of the corpus and formulated a new code book. Once all interviews were analyzed, the second pair drew comparisons from both code books and gathered frequently to resolve disagreements and identify important themes.

### 3.2 Findings

Participants regularly used tactile graphics, 3D models, and real objects for instruction. We outline their methods of acquiring and creating these materials, discuss educational and social aspects of use, explain their motivations for using one type of material over another, and describe their frustrations surrounding this material genre.

#### 3.2.1 Acquiring and Creating Tactile Materials

The process of acquiring tactile materials began in two ways. A general education teacher would request a tactile material for an upcoming class, or a TVI would ask a general education teacher what they plan to teach in upcoming weeks. Then, the TVI would use their past experiences with the student, and knowledge of the curriculum to prepare the materials that would be useful. There was no standard process for creating and acquiring tactile materials as it was highly dependent on the student's understanding and perception of the world, the participant's past experience with conveying the educational concept, and the materials available at hand.

Six participants specifically stated the importance of collaboration when finding the right material for the student. They often consulted and worked with the students' classroom teachers, paraeducators, therapists, and technologists (for assistive technology) to provide the right support. Elizabeth explained the significance of diverse perspectives, "If there's a vision issue, then I become very involved. If they are also seeing a physical therapist and occupational therapist, then they are very involved. We are all working together to see what their needs are." Three participants emphasized the importance of an online social network. Two participants used YouTube and Pinterest to gather ideas for tactile material preparation and one participant spoke about forming an online professional network through Facebook groups to share ideas and concerns regarding tactile materials with other TVIs.

Once the requirements were identified, participants would decide whether to purchase, use existing, or create new tactile materials. To purchase materials, 19 participants relied on specially designed tactile materials from the American Printing House for the Blind (APH), a non-profit that produces educational materials for students with visual impairments.

> *APH is our best resource. They've got a catalog of over 800 products so it kind of tailors upon what is needed for that student. I do have the tactile science notebook from APH that provides some diagrams. And it's just a question of whether or not, "Hey, do they align with the curriculum of what the student is using?" Sometimes they do. Sometimes they don't, and then it's up to me to create myself. (David, male, age 34)*

When participants decided to create their own tactile materials, they mainly created tactile graphics; only one participant spoke about creating a 3D model. Creating tactile graphics "is truly an art," Ashley explained, "you have to come at it with the attitude of, 'I'm making something that has to be understandable to somebody's fingers rather than their eyes.'" To create tactile graphics, participants used professional tactile material creation tools, including swell paper, Pictures-in-a-Flash (PIAF) machines, Perkins Braille Writers, and tactile material creation kits. Participants also used arts and craft materials, such as puffy paint, glue, clay, and Wikki-Stix, to create tactile graphics and 3D models. Participants described going shopping at dollar stores and thrift shops to look for materials, praising their affordable and diverse array of products. In fact, Connie proclaimed herself "the queen of the dollar store." "I think every vision teacher is," she explained. They often had to be creative with the types of materials they re-purpose.

> *I use a lot of things that some people might think are garbage. I have a whole bag of stuff, odds, and ends. And then I use the ties that are used on the breads, I'll twist that up and Wikki Stix, puffy paint. I use the bump dots, it depends, I use a lot of just tape or velcro. Anything that can feel tactile, just feel a difference from another surface. (Olivia, female, age 58)*

Interestingly, very few participants relied on technology to ease their creation process. Two participants used the PIAF machine, and three participants used braille embossers. In contrast, nine participants used tactile material kits which consisted of simple arts and crafts supplies. Tiffany noted that it was, "a very archaic way to do it, but that [was] the status quo."

Two participants emphasized the importance of involving students in the tactile material preparation process, explaining how creating these materials became lessons in themselves. Elizabeth's student needed a tactile map to navigate the hallways of his new school, and after many unsuccessful attempts with the braille embosser and raised line graphics, she used the Wheatley, a tactile diagramming kit from APH, and sat down with her student to create the map. "Me making the map was not working for him. He needed to be involved in making the map," she explained. "The tactile kit was perfect for that, so we sat down, and we made a map together. It took us several days to get it how we wanted it." The student felt confident about the map he helped to create and felt comfortable to rely on the map to navigate the school independently. Reflecting on this incident, Elizabeth said that most children did not need TVIs to take them to different places; rather, they needed a tactile material kit to create a map they could rely on and that is customized to their interpretation of the world.

### *3.2.2 Teaching with Tactile Materials*

Participants described two methods of instruction: push-in sessions and pull-out sessions. In a push-in session, the TVI accompanied the student into a general education classroom and supported them through a lesson taught by a general education teacher, whereas in a pull-out session, the TVI interacted with the student one-on-one in a separate session outside the general classroom.

Push-in sessions mostly occurred when TVIs conveyed concepts from subjects such as Math, Science, Geography and English. Alexis, a TVI who mostly presented tactile materials through push-in sessions, stated that it was "very difficult for classroom teachers [to have] a visually impaired student in their classroom." In her experience, classroom teachers would rely on the TVI to teach the material, whereas Alexis clarified that her duty was to, "adapt materials or teach specialized skills like Braille and Nemeth (math braille) to students so that they can participate in the regular-ed classroom".

In contrast, pull-out sessions were most often arranged when teaching braille literacy, an independent life skill, or other lessons that were part of the student's Individual Education Plan (IEP) or the Expanded Core Curriculum (ECC, see Section 2.3). When asked about their duties, 19 out of 21 TVIs mentioned teaching the ECC. "[I am] trying to work with them on recreation and leisure, teaching them to use the local transit, being able to understand how to get a taxi, making sure they can use money and not have it swiped, [...], functions of living, being able to follow a recipe and cook," Elizabeth explained, "there's all this stuff I do besides just academics." Tahani spoke about the importance of 3D models and real objects to teach the ECC, noting that it helped students to understand "what's going on in the world."

During the lesson, participants tried to engage several of the student's senses, including their tactile, visual (if student is low vision), auditory, and even their olfactory sense. "I have to give my blind students sensory time, like markers that smell, or oranges," elaborated Connie, "or for my deafblind student, I get her scents, so I'll have orange essence, or I'll cut up a lemon." Olivia explained how significant it was to incorporate tactile materials into the lesson in addition to providing verbal descriptions. "If they're hearing something and then you're giving them something to touch—it's a whole totally different concept to them," she said.

David found creative ways to incorporate multi-sensory feedback directly into a tactile material. He recalled a time he collaborated with his student's science teacher to teach animal anatomy with frog dissections. He borrowed a 3D model of a stomach from the doctor's office and added textures that resembled the insides of a frog's stomach. During the lesson, he also played stomach grumbling sounds.

Another use for tactile materials was to allow students to perceive their own creations. While sighted students receive immediate visual feedback during many tasks like drawing and writing, students with vision impairments relied on a TVI to indicate progress. To mediate this, Tahani used the PIAF machine to give her students tactile feedback from their drawings:

> *One of the things in my experience is that these students don't know what they're drawing. And then I'll actually take their artwork, and I'll tactile it, so they can experience what they just drew. It's actually been pretty funny. My kindergartner looked at me, she turned towards me, and she goes, "This is a mess. What is this?" And I just, I had to hold back the giggles, and I said, "No, this is what you drew." And she goes, "Well, it's a mess." [laughter] (Tahani, female, age 51)*

Tahani explained that when her students created artwork, it was a much more powerful and resonating experience for them to *feel* their drawings instead of just hearing a description of the drawings.

> *I think because there's a lot of self-discovery when you put a piece of paper in front of somebody, and most of the world, for my vision students, somebody else is telling them what they're seeing. Whereas when you tactile it, they actually get to find it out for themselves and read with their own fingers. (Tahani, female, age 51)*

To summarize, tactile materials allowed students to experience concepts on their own, which enabled them to be more independent. However, Maria explained that tactile materials were often too complex and abstract for students with additional intellectual disabilities.

### *3.2.3 Considerations for Different Tactile Material Types*

Identifying the appropriate tactile materials for a lesson is a skill that participants have honed through their years of experience. As an overview, Chidi shared her process for deciding on a tactile material:

> *It depends on the kid, it depends on how complicated the subject is, and it depends on what we readily have available and if we need to come up with something else. Often verbal description goes without saying, that's going to happen, and then my second choice would be 3D model or real object, followed by a tactile graphic. (Chidi, female, age 31)*

Participants considered several factors before finalizing the materials for a lesson: (i) the creation time at hand, (ii) the student's mode of learning, (iii) the availability of real objects, and (iv) the independence offered by the tactile material.

*Creation Time at Hand*

Time was the largest factor for TVIs when deciding which tactile materials to use. The lack of time often hindered participants from obtaining the materials they would have preferred to work with. Blossom and April spoke about how creation time largely dictated the kinds of materials they could present to the student.

> *The ultimate decider is time. If there is a situation where we haven't been given enough advance notice to make a graphic, or can't make a 3D model, then we'll probably do a verbal description. But I feel with those two students that I use tactile things with every day, that there's more concepts [where] they have something to put their hands on rather than just something you told them. (Blossom, female, age 37)*

Creating 3D models was especially time-consuming. April would "create more of the graphics than the 3D model. The only reason why is time." Participants wished to present 3D models to their students but were unable to do so because of the time and effort required to obtain them. To prioritize, Lauren would consider "how important [the concept being taught] is to the curriculum." She would spend more time to create 3D models for concepts that were fundamental to the student's learning.

*Students' Mode of Learning*

Tactile materials were very specific to a student's needs, preferences, and mode of learning (i.e., auditory or tactile). Participants explained that when a new student was assigned to them, they spent several lessons to understand how the student learned and grasped concepts. Four participants mentioned that their students were "tactile defensive," meaning that the student had a strong aversion to certain materials. One of Connie's students refrained from touching braille because the feeling of the dots disturbed her, similar to how students with vision may find a sight disgusting to view. For tactile learners, Olivia explained "[their] first choice might be the tactile graphic and then for other students it might be the last resort."

According to Elizabeth, classroom teachers frequently resorted to explaining concepts auditorily to her students, which did not align with their mode of learning. "Although descriptions and stuff like that are wonderful with auditory, it would be really nice to always have something under their fingers because not all of my students are auditory learners," she elaborated. Some of her students were more "kinesthetic learners", and some "low vision students [were] visual learners which is kind of funny." Similar to Elizabeth, other participants also suggested that many of their students were tactile learners, but that they needed additional auditory or kinetic components.

*Availability of Real Objects*

When possible, participants used real objects to explain concepts and provide students with an accurate representation of the world. Seven TVIs preferred to teach with real objects as opposed to tactile graphics or 3D models. Alexis remarked that tactile graphics made "very little sense" to her students if they have not encountered a real object of the concept previously.

> *I've gotten samples where they'll say tire, and they'll have the rubber. It's a hard concept to comprehend when you think of a tire, yeah, it's made of rubber, but that's not really what a tire is, it's not a little piece of rubber. I've taken kids outside and actually walked around a car and had them feel a car and feel the tire and try to get an idea of the scale. (Olivia, female, age 58)*

Several participants explained that 3D models could not adequately represent the accuracy, size, and texture of real objects. Alexis stated that 3D models actually had potential to confuse students because of these inaccuracies in the representations. Elizabeth and Alexis both used the example of a bus to illustrate this point.

> *Sometimes I would say, like if you're not using an actual object, you can kind of confuse a student. Like a bus is a good example. A kid gets on a bus, but you really can't, for safety reasons you might not be able to use an actual 3D representation... you could use a representation of a bus, but that doesn't give you the information about the actual size of a bus. And if you want that information, then you have to get out and touch a bus. (Alexis, female, age 50)*

Elizabeth expanded on this perspective, explaining that a 3D model bus, "feels like a block that's a square with something round on the end. That's not going to make any sense to them." Similar to Alexis, she wanted the student to, "actually feel and touch [and] walk through and feel the outside" of a real bus. However, feeling a real object was not feasible for all concepts. Alexis gave the example of a building, "even if you feel the building, you're not feeling the height of the building, you're not getting an idea of how tall it is." So although many participants turned to real objects to convey different textures and scales, they were not able to acquire and present the actual referent objects if they were too large or dangerous.

*Independence*

Thirteen participants suggested that tactile materials provided social benefits and promoted independence. During push-in sessions, when students with vision impairments learned at the same pace as other sighted students, participants were mindful to present materials in a way that did not require them to provide verbal descriptions. Rebecca contacted the classroom teacher in advance, gathered information about the curriculum, created the tactile materials, and pre-empted her students with the materials they should expect. "That way when they're in class and they encounter [the tactile material], it's not something new, they don't need an assistant in the classroom to say, 'Oh, that's what this symbol is, and this is how you make it,'" she explained. She steered away from one-on-one sessions because her students "are not going to have a one-on-one in adult life."

Interestingly, three participants noted that 3D models promoted more independence than tactile graphics when learning concepts. According to Donna, 3D models did not require as much supervision:

> *The advantage of using the models is that kids can experience it at their level, firsthand. They can make judgements about what they're touching. They can interpret the relationship by using both hands together with a model. The tactile graphics, the pictures and other things that are not 3D, I think that it really limits* *the child. There can be too many details and the kid just can't interpret it well and by themselves. They need constant, I would say, supervision or guidance. (Donna, female, age 58)*

Tahani justified the benefits of 3D models with a different nuance. She used a 3D model of a bird to explain how students drew connections and learned independently once they had explored the model:

*They can turn it around in their hands completely and feel the bird wings and feel feathers, they just seem to connect more with it. As they have a foundational feeling of what a bird was, then when I would tactile a bird on a page, it was a faster connection for them. And it seemed to promote more independence as they moved from the 3D model then to the actual tactile page. They conceptually understood in their brains. If they had never seen a bird, they had felt what a bird was, and then when it moved to the page, they could find the wings. They could find the beak, because they had touched it. (Tahani, female, age 51)*

Despite these positive impressions, participants identified some issues that hindered their use of 3D models in the classroom. While tactile graphics closely resembled the materials used by sighted students, 3D models were obtrusive and instigated feelings of exclusion among their students. Two participants recalled that their students needed help moving their models around, which negated the independence the models could afford.

### *3.2.4 Challenges and Frustrations with Tactile Materials*

Participants identified several frustrations with tactile materials.

*Poorly Designed Tactile Graphics*

Many participants found that professional tactile graphics were poorly designed—they contained too much information and lacked tactile distinction. Rebecca succinctly explained, "for a sighted person to take a two-dimensional picture and make it a three-dimensional object in [their] head is hard enough. For a blind student, obviously we're looking at nearly impossible." Yet, several participants encountered tactile graphics that merely transferred abstract concepts or 3D representations to a 2D tactile graphic.

Ashley spoke about the tactile graphics in her students' biology textbooks from APH, "there's a lot of good detail," but, "if you provide too much tactile information while a child is trying to explore, it actually makes less sense in their brain." Embossed tactile graphics, such as the ones in these textbooks, usually lacked tactile contrast. In fact, this was a large concern regarding braille embossers. According to participants, braille embossers produced dots of equal size and height. This was frustrating and confusing for students as they tried to distinguish between different elements of a crowded tactile graphic. While there are modern embossers available that provide variable height, Tahani, like most participants, used a tactile material creation kit to combat this issue.

*[In the embossed picture], all the dots are exactly the same. But if you want a child to differentiate between what's happening on a tree, the bark, the leaves, the stems, I can do that with the tactile kit, whereas on the embosser, it just all comes out in the same dots. So, there's no break in that tactile feel. And I like giving that much detail to them, to say, "This is what the tree feels like," and having the lines going down like bark strips, and then using a different tactile tool to outline the leaf. And it just gives them a better sense of what the tree feels like. (Tahani, female, age 51)*

Although some tactile graphics had braille labels, participants shared that not all students were able to read braille. Participants further criticized braille labels, explaining that they were poorly placed, took up too much room, or were interpreted as a texture. In general, TVIs preferred simple materials which were easily comprehensible and highlighted the salient features of an image.

*Tactile Materials for Students with Low Vision*

Six participants spoke about the challenge of finding tactile materials for students with low vision. According to them, most tactile materials were designed for those with no vision and heavily relied on braille labels. "Braille labels are good for the kids who are braille readers," Tiffany explained, "we have 100 some, approximately, visually impaired kids in our school district. Of those, maybe 10% are braille readers. The other 90% are reading print." Three other participants also noted that the majority of students they work with have some residual vision, allowing them to see high contrasts, bright colours, and even large print.

> *It was just a shame for me to have this wonderful 3D model that doesn't have color, doesn't have other markings on it. If I'm really trying to teach the difference between a zebra and a horse and I've got a kid who's got very, very poor low vision, I need those stripes. Why not have them? To me, it would just be a shame to have developed everything you've developed and not have it be available to the majority of the visually impaired kids who are out there. The majority of them can see something. (Tiffany, female, age 64)*

Eight participants spoke about having to specially create or adapt tactile graphics for students with low vision. Christine carried, "highlighters, fluorescent duct tape, low vision markers, […] Ticonderoga pencils or carpenter pencils," to the classroom everyday to adapt visual and tactile graphics to be accessible to her students with low vision. Kamilah praised a tactile diagramming kit from APH as it contained bright *and* textured materials for her to use "with low vision and blind students."

*Tactile Materials for Students with Additional Disabilities*

All participants prepared materials for students with multiple disabilities. According to them, tactile materials were primarily designed for students that were "vanilla blind"—a term used by Rebecca and her colleagues to refer to students with only visual disabilities. These materials lacked considerations for students with intellectual disabilities, who needed greater sensory simulation, and for students with physical disabilities, who needed easy maneuverability. Having low vision herself, Rebecca spoke from the perspective of a student growing up in the special education system, and from the perspective of a TVI.

> *I was visually impaired growing up so my last four years of high school I went to a school for the blind myself. What I have seen, knowing the population of those schools back in '90 versus the population now, the number of 'vanilla blind' children is almost non-existent now. Very rarely do we find a vanilla blind child. In a caseload of thirty, I had two who were not multiply disabled. There's at least a learning disability or sensory issues. (Rebecca, female, age 40)*

Participants often used real objects for sensory stimulation or employed tactile material creation kits to make materials extremely personal to student with intellectual disabilities. Olivia played patriotic music to engage her student with Cortical Vision Impairment, a brain-based visual impairment, during her art project. Other participants spoke about the challenge of presenting materials to students with physical disabilities. Tiffany disclosed that, "fifty percent of [her] kids are in wheelchairs," and she often thought of ways to allow her students with physical impairments to manipulate tactile materials without needing to utilize both hands. "A lot of times, I'll take my kids and I'll grab a can of playdough," she explained, "I'll stick the blob of playdough on the desk. Then, I would take a cow model and I would stick his legs in the playdough." This allowed her students to free their hands in order to balance or regain control and use one hand to explore the model. Two other participants faced similar challenges when introducing 3D models to their students with additional physical disabilities.

### 3.3 Non-Tactile Accommodations

In addition to tactile materials, participants also mentioned a variety of accommodations they provided for their students. This involved verbally describing visual materials and concepts, physically guiding the students' hands, and using their body as a learning tool. Some participants used assistive technology such as tablets, refreshable braille displays, magnifying software or devices, and software or devices with speech output.

### 3.4 Discussion: Key Findings and Open Questions

Our Interview Study offered new insights that underscored the growing discrepancies between current innovations in tactile materials and current practice, provoking many additional questions. In this section, we highlight key findings relative to prior research, and pose questions for deeper inquiry.

While technologists have often seen tactile materials as a way to replace visual imagery [32,42,61], we found that tactile materials are a distinct medium that require prerequisites such as tactile literacy and spatial awareness. All participants used tactile materials to help students build these foundational concepts, relying on real objects and 3D models as important teaching aids. Previous research has focussed on understanding the use of tactile graphics [2,45,69–71], but our investigation of all types of tactile materials revealed that teachers preferred to use real objects and 3D models before introducing their students to tactile graphics. However, real objects were difficult to acquire for large or abstract concepts (Section 3.2.3), and many participants were unable to recall the last time they created a 3D model. Although there has been much recent work that leverages 3D printing for tactile models [50,52], none of our participants used 3D printers for this purpose.

Thus, our understanding of the use of 3D models is still limited. How do TVIs obtain 3D models and what affordances do they offer beyond those of tactile graphics? Additionally, how are 3D models used to build foundational skills (e.g., tactile literacy), and how do TVIs incorporate these materials into their lessons?

Another key finding was that tactile materials were indeed time-consuming and laborious to prepare, as noted in prior work [48,69,70]. Beyond prior work, our study shed light on the range of materials used by TVIs and the decisions they made during their creation process. Participants collaborated with other staff to identify appropriate materials for their students, went shopping for general materials at dollar stores (Section 3.2.1), and collaged unconventional materials together to create tactile materials. However, these findings were fairly general and could benefit from concrete examples, which have not yet been presented in literature. How do TVIs repurpose general materials as tactile materials, and what do these tactile materials entail? Answers to these questions will help technologists who seek to design tactile material creation tools.

Participants also expressed concerns regarding tactile materials that were only designed for students with no vision, including materials that assumed all students with vision impairments read braille. Prior work has addressed and mitigated this challenge with audio annotations [4,52], but several participants spoke about the importance of letting students "self-discover" concepts through tactile materials. Tactile materials, especially 3D models, allowed students to gather information on their own, without relying on verbal descriptions. How do TVIs facilitate this learning process? Do they design tactile materials in ways that did not require verbal explanations? Participants further spoke about customizing existing tactile materials to meet the needs of students with low vision or additional physical and intellectual disabilities. How exactly do they modify or enhance these materials to suit such a diverse range of abilities?

Prior work suggested that tactile materials supported classroom inclusion [71], and we reveal the considerations regarding this benefit. Despite the independence that 3D models offered, they were often obtrusive, unlike tactile graphics that closely resembled printed materials used by sighted students. This prompted feelings of exclusion among students with vision impairments and hence, participants preferred to use tactile graphics during push-in sessions. This raised larger questions around how teachers balanced the use of tactile graphics and 3D models to foster self-reliance within students, while also allowing them to feel at pace with their peers.

Although the Interview Study addressed our research questions, it also fueled the need for a deeper understanding of our findings. We wished to know more about how TVIs use tactile materials to teach non-academic concepts from the ECC (i.e., tactile literacy, spatial awareness), and how they personalized tactile materials for their students. We wanted to better understand the day-to-day decisions that TVIs made regarding their use of tactile materials. Therefore, we conducted a diary study, where participants provided information about specific tactile materials they used with each of their students. We discuss the methodology and findings of our diary study in the subsequent sections.

## 4 DIARY STUDY: EXAMINING TACTILE MATERIALS IN PRACTICE

### 4.1 Methods

We conducted a diary study to gain a comprehensive understanding of tactile material use.

*Participants*. Of the 21 participants in the *Interview Study*, eight partook in the diary study (see Table 1, designated with asterisks). All diary study participants were female, and their ages ranged from 31 to 58 years with a mean of 43.8 years (SD = 10.6). Two participants identified themselves as blind, while the rest identified as sighted. Their professional experience as a TVI or braillist ranged from one year to 24 years. Participants were financially compensated for the diary study.

*Procedure*. A diary entry was structured as an online form, and participants were instructed and guided to save a link to this form on their smartphone, tablet, or computer. Participants were asked to complete a diary entry for each tactile material they used over a 3-week period. Aligning with the intent of the study, participants were not given explanations or definitions of what constitutes a tactile material, or what should be reported; they were simply encouraged to report any and all "tactile materials" they used with their students. First, participants were asked to attach at least three photos of the tactile materials. Then, participants were asked the following eight questions about the material:

- What concept(s) in the curriculum were you helping your student(s) with when using the tactile material?
- Regarding the student(s), what grade(s) are they in? What disabilities do they have?
- What other kinds of accommodations did you consider, if any?
- Why did you choose that tactile material over other accommodations?
- If you purchased/received the tactile material, where/whom did you purchase/receive it from?
- If you created/adapted the tactile material, what materials did you use? In particular, which technologies or low-tech methods did you employ?
- Approximately how long did you spend in total to acquire, create, and/or adapt the tactile material?
- Approximately how many USD did you pay to acquire the tactile material?

- How would you improve this material if you had additional time and/or resources?

At the end of the three-week period, the diary study concluded with a final interview via phone or video conference software; it was recorded and lasted approximately 15-30 minutes. We asked participants to clarify any ambiguous or unanswered diary responses and provided them with an opportunity to add any additional comments regarding their experiences with their students and tactile materials.

*Data and Analysis*. We collected a total of 70 entries, with each participant submitting 9 entries on average. Each entry included at least three images and short responses to open-ended questions. These entries were analyzed by two researchers. The images were indexed, catalogued, and categorized based on attributes of the tactile material such as its size, its composition, and other characteristics. The short responses were analyzed using a mix of open and closed coding methods. Open codes were used to document the creative process of tactile material preparation and to understand the participant's decision-making process when considering alternate materials. Closed codes were used to categorize the entries based on groupings such as type of tactile material and the concepts they taught. Based on the standard types of tactile materials in practice, the materials were placed into one or more of the following types: (1) Tactile Graphics, (2) 3D Models, (3) Real Objects, and (4) Braille Embossed. The concepts conveyed through these tactile materials were grouped into the following areas: (1) Braille Literacy and/or Reading/Writing Skills (2) Math (3) Orientation and Mobility Skills (4) Social Skills and Daily Living (5) Art (6) Fine Motor Skills, and (7) Science. These categories were derived from subjects in the standard curriculum, as well as the Expanded Core Curriculum (ECC) (Section 3.2.2). Braille Literacy and Reading/Writing Skills included subtopics such as braille code, storytelling, word processing, comprehension, and History. Math included subcategories such as shape and pattern matching, Geometry, and Economics. Orientation and Mobility Skills (O&M) included subcategories such as map reading, gross motor movement, intersections, pedestrian patterns, tactile scanning, and cardinal directions.

In addition to the codes, one researcher detailed their thoughts and interpretations through analytic memos. These jottings were discussed with other researchers to examine the diary entries in depth. To draw relations between the interviews and diary entries, researchers used the final code book from the interview analysis to code all the diary entries.

### 4.2 Findings

In this section, we present findings from our analysis of the diary entries. All quotes and images were directly taken from the participants' diary entries.

#### *4.2.1 Overview*

Table 2 shows an overview of the 70 diary entries collected from the 8 participants. These ranged from 5-14 entries per participant with a mean of 8.75 (SD = 2.86). Tactile materials were presented to students from Pre-K to 10th grade, with an average of 3 different grades represented per participant. The students' visual characteristics were identified as "low vision," "blind," and/or "cortical visual impairment (CVI)."

Six participants submitted tactile materials for students with intellectual disabilities. This constituted 26 of the 70 entries (33.3%). The intellectual disabilities identified, in participants' own words, were autism, learning disabilities, communication disabilities, attention disabilities, and developmental delays. Two participants presented tactile materials to students with physical disabilities, including students who were hard of hearing or had mobility challenges, which amounted to five out of the 70 entries (7.1%).

Table 2: An overview of the diary study entries, collected over the course of 3 weeks.

| Participant | No. of Entries | Tactile Graphics (%) | 3D Models (%) | Real Objects (%) | Braille (%) | Student Grade | Topics Covered |
|---|---|---|---|---|---|---|---|
| Tahani | 10 | 2 (20) | 7 (70) | 4 (40) | 1 (10) | Pre-school, K, 6 | Braille Literacy, Reading/Writing, Math, O&M |
| Eleanor | 8 | 5 (63) | 2 (25) | 3 (38) | 3 (38) | 6, 7, 9 | Braille Literacy, Reading/Writing, Math, Geography |
| Blossom | 5 | 1 (20) | 2 (40) | 2 (40) | 1 (20) | Pre-school, 4, 5 | Math, O&M |
| Chidi | 6 | 4 (67) | 1 (17) | 1 (17) | 4 (67) | Pre-school, 2, 4 | Braille Literacy, Reading/Writing, Math |
| Alexis | 10 | 9 (90) | 3 (30) | 0 (0) | 1 (10) | 1 | Braille Literacy, Reading/Writing, Math, Geography, History, Art |
| Olivia | 11 | 3 (27) | 2 (18) | 3 (27) | 5 (45) | Pre-school, K, 2, 3, 10 | Braille Literacy, Reading/Writing, O&M, Social Skills & Daily Living, Fine Motor Skills |
| Connie | 14 | 8 (57) | 5 (36) | 4 (29) | 2 (14) | K, 6 | Braille Literacy, Reading/Writing, Math, O&M, Art, Fine Motor Skills, Daily Living |
| Kamilah | 6 | 4 (67) | 4 (67) | 1 (17) | 0 (0) | 2, 3, 4 | Math, O&M, Daily Living |
| Total | 70 | 36 (51) | 26 (37) | 18 (26) | 17 (24) | – | – |

Note: Categories of materials were not mutually exclusive. i.e., an entry could contribute to the tactile graphic count, and the 3D model count, if both types were used.

Of the 70 diary entries, 36 (51.4%) involved tactile graphics, 26 (37.1%) involved 3D models, 18 (25.7%) involved real objects, and 17 (24.3%) involved braille, as embossed or labeled. As evident with these values, entries fit into more than one category. The most common combinations were that of tactile graphics with 3D Models (22 entries, 33.3%), and tactile graphics with braille (10 entries, 29.4%).

Participants specified the educational concepts they aimed to teach while using the tactile materials. These subjects included: (1) Braille Literacy and/or Reading/Writing Skills, 31.4%; (2) Math, 23.8%; (3) Orientation and Mobility Skills, 15.8%; (4) Social Skills and Daily Living, 11.9%; (5) Art, 8.0%; (6) Fine Motor Skills, 5.1%; and (7) Science, 4.0%.

### 4.2.2 Acquiring and Creating Tactile Materials

The diary entries revealed how participants financed their tactile materials and shed light into their creative process. Table 3 reveals a taxonomy of the TVI tactile material acquisition process. We tally the number of times TVIs chose to create or purchase a tactile material, and whether they adapted any of these materials for their student's abilities and habits. We also include an average time and cost incurred by TVIs, as disclosed through their diary entries.

*Purchasing Tactile Materials*

Over half of the tactile materials reported (52.8%) were obtained by independently ordering and/or purchasing through online resources. The most common online resource was APH, seven out of the eight participants purchased tactile materials from APH, amounting to 24 (35.3%) of the total entries.

All participants relied on quota funds to cover the financial costs of the tactile materials; these funds are given to every school district by the Department of Education, Office of Special Education Programs, and are allocated based on the number of eligible students with vision impairments that each district supports. They are designated for purchases of educational materials from APH. More than half of the entries (40 entries, 57.1%) reported no out-of-pocket expense for the participant, implying that materials were available at school, or were purchased using quota funds. When participants were not using quota funds, they used free resources from local organizations, such as the New York State Resource Center. Despite the allocated funds, all participants personally expensed at least one tactile material, most participants only spent $5 or less (15 out of 70 entries, 21.4%), some participants spent between $10 and $30 (4 entries, 5.7%).

Table 3: A breakdown of tactile material acquisition by each TVI, over the course of 3 weeks.

| *Participant* | *No. of Entries* | *Created (%)* | *Purchased (%)* | *Adapted (%)* | *Avg. Time Spent\* (mins)* | *Avg. Cost Incurred\* ($)* | *Materials & Technologies Used* |
|---|---|---|---|---|---|---|---|
| *Tahani* | 10 | 4 (40) | 6 (60) | 3 (30) | 49 | 5 | Cardboard, Beads, Strings, LEGO Blocks, Puzzles, Perkins Braille Writer, Laminator |
| *Eleanor* | 8 | 6 (75) | 2 (25) | 2 (25) | 30 | 3 | Cardboard, Foam Dots, Pipe Cleaners, Abacus, Braille Peg Slate, Raised Ruler |
| *Blossom* | 5 | 4 (80) | 1 (20) | 2 (40) | 9 | 1 | Velcro, Beads, Jar, Magnets |
| *Chidi* | 6 | 5 (83) | 1 (17) | 2 (33) | 55 | 6 | Braille Paper, Toys, Braille Embosser, Abacus, Puffy Paint, Foam, Feathers |
| *Alexis* | 10 | 6 (60) | 4 (40) | 7 (70) | 51 | 5 | Rubber Bands, Pegboard, Wikki Stix, Perkins Braille Writer, PIAF Machine |
| *Olivia* | 11 | 3 (28) | 8 (72) | 3 (28) | 5 | 2 | Clothes, Braille Peg Slate, Velcro, Felt, Beads, Quick-Draw Paper, Textured Paper |
| *Connie* | 14 | 3 (21) | 11 (79) | 2 (14) | 2 | 2 | Wikki Stix, Tactile Drawing Board & Film, Paint, Braille Swing Cell, Foam Shapes |
| *Kamilah* | 6 | 2 (33) | 4 (67) | 3 (50) | 2 | 2 | MathLink Cubes, Traffic Cones, APH Tactile Town Kit, Felt Sheets, APH Picture Maker |
| *Total* | 70 | 33 (47) | 37 (53) | 24 (34) | 25 | 3 | – |

Note: Created materials include materials that were found and assembled for the lesson. Adapted materials refer to tactile materials that were found, created, or purchased, and *then* modified for the student's needs and abilities. *These values were reported directly by the participants. In some cases, the department funded the item and participants incurred a $0 cost or spent 0 mins to browse the material. These values are included in the avg. calculations.

*Creating and Adapting Tactile Materials*

All participants submitted tactile materials that they created themselves, which amounted to 33 of the 70 entries (47.1%). Less than half (43.4%) of these materials were created using tools designed for tactile material production (i.e., swell paper, PIAF machines, Perkins Braille Writer), whereas 56.6% of the materials were made using general arts and crafts materials (i.e., tape, clay, puff paint). A greater variety of technologies and materials are described in Table 3. The entries demonstrated (i) why participants chose to create or adapt existing tactile graphics, and (ii) how they combined tactile graphics, 3D models, and real objects to deliver a concept.

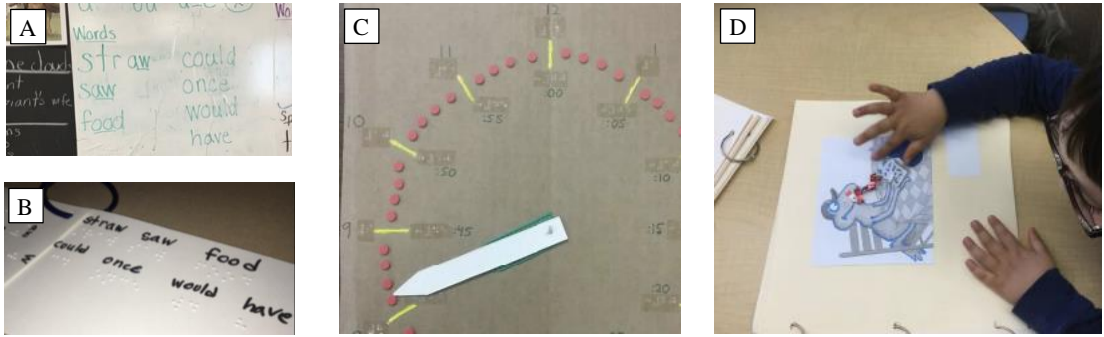

Figure 1: Materials for push-in and pull-out sessions. (A) Whiteboard in Alexis' student's English class. (B) Binded index cards created by Alexis to communicate the words from Figure 1A. (C) Tactile clock by Eleanor, materials used to create this clock included, part of a cardboard box, puff paint, adhesive braille paper, a permanent marker, foam dots, a protractor, a twist tie, plastic craft canvas, and parts of a cardstock material box. (D) A tactile book created by Chidi, pictures are outlined with puffy paint and contain foam pieces.

The entries indicated a stark difference between when participants simply chose to translate visual material (most often into a braille description), and when they took the time to find or create the appropriate tactile material to convey an abstract concept (most often involving tactile graphics or 3D models). During her push-in session, Alexis created a braille embossed booklet for her student which directly translated the words on the board to braille (Figure 1A). There was no conceptual barrier that Alexis needed to clarify for the student. In contrast, Eleanor and Chidi created tactile materials and *adapted* them to address and fill specific gaps in their students' conceptual understanding. Eleanor created a tactile clock (Figure 1C) to explain the concept of reading time. The readily available tactile clock from APH was insufficient because she, "needed a larger clock to distinguish the minute marks easier, and to have the 5-minute markings brailled." Eleanor spent a total of three hours to create that larger, more meaningful clock, while using materials found in her classroom and at her home. Similarly, Chidi created her own tactile graphics. She used art supplies to emphasize significant features of a story illustration to her student, (Figure 1D), using "puffy paint to outline frog pieces of foam and feathers to make the bugs and other items." This, as opposed to a simple braille description, enhanced the pictures and "gave meaning to the story." The emphasis of the 5-minute marks by Eleanor and the prominent textures in the story book by Chidi provided more than translations of the curriculum.

Participants did not make distinctions between different types of tactile materials (e.g., tactile graphics vs. 3D models) and often combined them to create one tactile material. In an ideal scenario, they preferred to present tactile graphics *with* 3D models. Tahani described 3D models as a "stepping stone to tactile graphics," and Blossom explained that without a concrete 3D representation of the concept, tactile graphics may be "completely meaningless." These mixed tactile materials were created and used by all eight participants, (40 entries, 57.1%) to (i) indicate the salient parts of a lesson, and (ii) support the student's understanding of a concept.

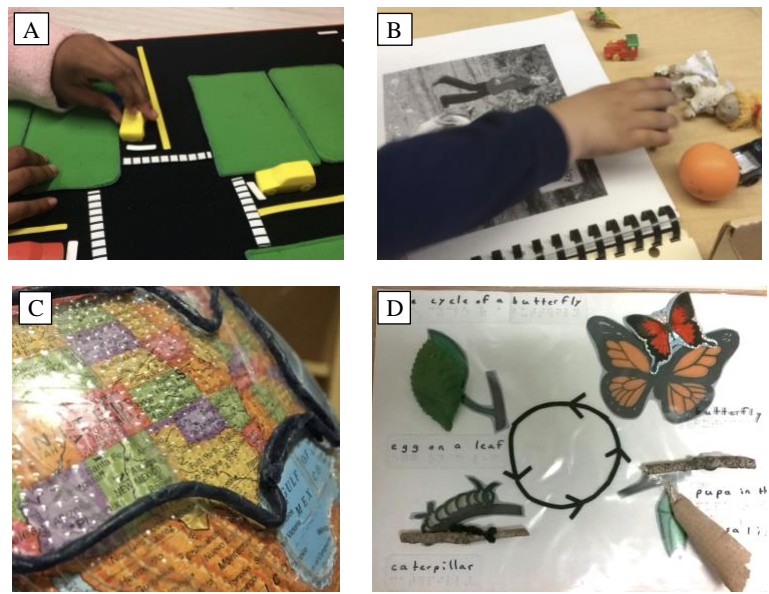

Figure 2: Mixed tactile materials. (A) Kamilah's student interacts with tactile roads, lanes, and cars to learn intersection traffic patterns (B) Chidi facilitated his student's reading session with 3D models and real objects. (C) Tactile Globe from APH, adapted with Wikki Sticks by Alexis to bring focus to the US region. (D) Life cycle of a butterfly, created by Eleanor and a paraeducator, using graphing tape, pipe cleaners, fake leaf with hot glue dots (to indicate a butterfly egg), and paper towels.

To indicate the saliency of the car in street intersection patterns, Kamilah used 3D models for the cars, while using tactile graphics for the rest of the components (Figure 2A). The car could have been represented as a flat graphic, but the models provided depth and conveyed to the student that cars are more noteworthy and dynamic than the rest of the components. In another example, Chidi used 3D models along with the tactile graphics of a story book (Figure 2B). This helped her pre-braille student (not yet able to read braille, similar to pre-reading) to make, "connections between pictures and 3D models and enhance the story." Eleanor used a mixture of tactile graphics, 3D models, and real objects to explain the life cycle of a butterfly (Figure 2D). The multiple textures helped her student to discern between different cycle stages in a meaningful, resonating, and engaging way. Sometimes, participants added tactile graphics atop the 3D models to adapt them to their students' needs, such as Alexis' outline of the US map on a tactile globe (Figure 2C), to draw attention to certain parts of the model.

### *4.2.3   Teaching with Tactile Materials*

Most tactile materials were presented in pull-out sessions and covered a range of educational concepts. The most common topics covered were Braille literacy, Math, and O&M. Confirming our interview findings, all participants conveyed skills pertaining to the ECC. Eleanor taught her student to recognize tactile features of a dial pad and to dial a phone using a real telephone, and Blossom taught her student the motion of pouring from one container to another using jars and beads (Figure 3A).

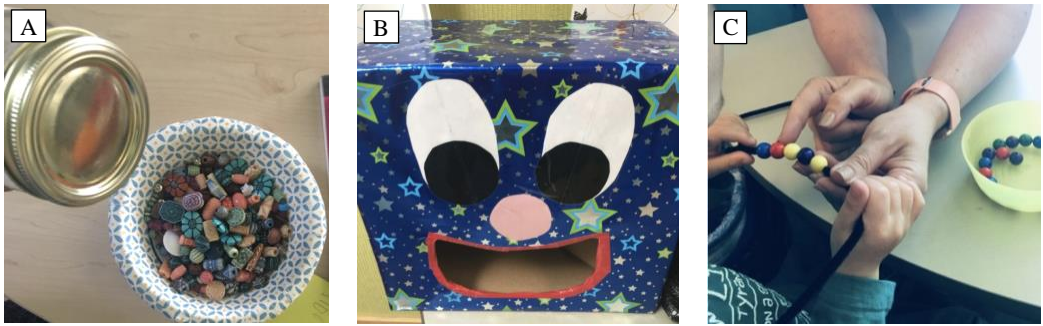

Figure 3: Teaching with tactile materials. (A) Blossom brought beads from her home and let her fourth grader practice unscrewing lids and pour from one container to another. She used beads instead of liquids or real food to avoid a "sticky mess" and to transition to another lesson with beads. (B) Braille Monster created by Tahani. Her students fed braille cards into the monster's mouth, and if the braille was incorrect, she made "buzzy noises." (C) Tahani co-created a bracelet with the student. In the image, the student strings the beads himself, as the teacher holds the beads in place.

Participants added elements of fun to their teaching practice. For her braille literacy class, Tahani purchased a cardboard box and adapted it to resemble a monster's face (Figure 3B). Students fed braille index cards into the monster's mouth, and Tahani made "buzzy sounds" if the answer was wrong. Tahani's main intent was "to make learning Braille fun," and present academic content in a way that was easily digestible. Similarly, for her session with pre-school students, she used textured wheels because her students found its rotations "entertaining."

Three participants involved their students in their tactile material preparation process which often became the lesson in itself; it challenged the students' fine motor skills and spatial sense. Tahani gave her student the materials to create his own bracelet (Figure 3C) because it supported the student's "independence in creating it, [increased] smaller motor skills, [and created] excitement about having it on his backpack." In other cases, the participant brought professionally produced tactile materials to the student, such as puzzle pieces and flip-over books (Figure 5). These materials often needed rearranging, so the student created their own scenarios and learned from them.

### 4.2.4 Challenges and Frustrations with Tactile Materials

More than half of the entries (36 out of 70, 51.7%) offered suggestions to improve existing tactile materials given additional time and resources. A majority of these improvements (15 of the 36, 41.7%) included increasing the tactile contrast of the materials, a concern also expressed by several participants in the *Interview Study*. Participants felt that existing tactile materials did not distinguish salient features of a tactile material and often lacked tactile contrast. Chidi wished that an embossed map he purchased had "different dot heights" and more spacing between the dots. She, along with other participants, also wished for additional tactile cues atop 3D models. Alexis purchased a globe from APH and found that it had a similar tactile feel across the surface. To combat this, she used Wikki Stix to outline the border along the United States (Figure 2C) but additionally wished for a globe with raised relief information to indicate different land masses (*i.e.,* mountainous regions).

The second largest concern (8 of the 36, 22.2%) was regarding the stability and durability of tactile materials, followed by the challenge of organizing and storing tactile materials. Four participants reported concerns of durability, explaining that they would use sturdier materials to improve the tactile material. Due to the deterioration of Eleanor's butterfly life cycle diagram (Figure 2D), portions have had to be remade over the years. Chidi reported a similar challenge with the tactile book that was not, "holding up too much wear." The challenge of creating durable and long-lasting materials often stemmed from the rapid speed with which teachers had to create these materials. For example, Olivia had less than 30 minutes to ideate, gather materials, and create a tactile direction chart for her student (Figure 4A-C). The lack of durable materials often made storing tactile materials very difficult. Participants had to find ways to organize and store tactile materials to ensure their integrity for future use; many participants laminated tactile graphics and maintained a storage room of tactile materials. Alexis used ring to bind materials together (Figure 1B), and emphasized that, "it's not just [about the] materials, it's how you keep them organized." Interestingly, Blossom encouraged her student to organize the components of a tactile material by allowing them to move the magnets and set up math problems themself (Figure 5D). Blossom used the Math Window instead of letting her student use a braille writer so they could, "practice organizing and moving items around to build spatial awareness."

### *4.2.5 Considerations for Different Tactile Material Types*

The largest decision-making factor for considering a tactile material was its prompt availability (24 entries, 37.1%). Materials were chosen because they were readily available or very easy to create, such as Olivia's directional chart (Figure 4A-C), confirming findings from the *Interview Study* that creation time can be a large hindrance (Section 3.2.3). The second factor was high student engagement levels associated with the materials (16 entries, 22.8%). Participants frequently selected materials because their students would find them fun, interactive, or exciting. Another consideration (16 entries, 22.8%) was reusability, participants preferred materials that would serve the learning goals of multiple lessons and continue to challenge the student. Lastly, although not explicitly pronounced, participants gravitated towards materials that encouraged movement, involved real objects, and fostered independence.

*Fun and Engaging Tactile Materials*

All participants emphasized the element of fun. The primary reason Tahani chose to create the Braille "monster box" (Figure 3B) or offer spinning tactile pattern matching wheels because of the "fun appeal." Tahani could have displayed patterns in books and tactile graphics, asking her students to match different textures, but she preferred 3D models instead because the rotation was, "entertaining for the kiddos," (Figure 5C) However, engagement was highly dependent on the student's interests and abilities. Connie brought bubble wrap for her student because he really enjoyed popping bubbles (Figure 4D), and she brought drums and gel balls for another student because he enjoyed the motions and textures associated with those items (Figure 4F). Although these tactile materials were not conveying standard academic concepts, recreation and leisure is an important aspect of the ECC, and we noted several tactile materials serving this purpose.

*Reusability*

Seven participants leaned towards materials that could be reused for multiple lessons, a notion we define as reusability. Used as a concept reinforcement tool, these tactile materials had varying levels of complexity and would continue to challenge the student beyond the first lesson. Tahani did not consider any other accommodations when she selected this shape sorting tactile material (Figure 5A) because it helped her to, "stretch the learning skills," of the student by introducing, "one [additional shape] at a time." Similarly, Olivia used the Line Paths book (Figure 5B) frequently with her student because she could facilitate multiple lessons and let her student, "flip cards and reassemble." Allowing a build-up of skills was the charm to these materials; many participants wished there were more objects, textures, and other complexities integrated within their material.

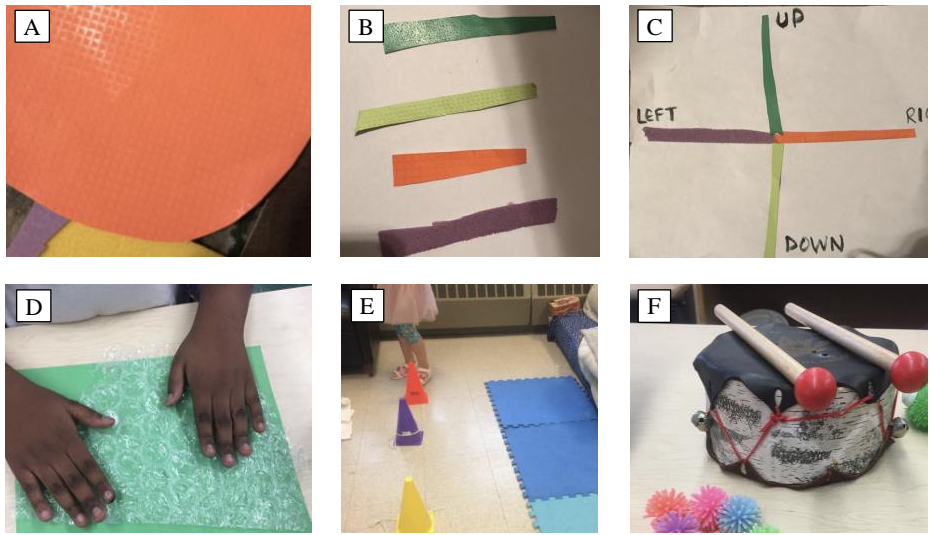

Figure 4: (A-C) Olivia's rapid process of creating a tactile material to explain directions to her pre-school student. (D) Bubble wrap, collected by Connie to present to her student because he, "gets great joy out of popping the air out of the bubbles!" (E) Kamilah used traffic cones and foam mats to guide her student through traffic and intersection patterns. (F) Items from Connie's Sensory Box, drum covered in rubber with jingles, drumsticks, and balls filled with gel.

*Movement and Incidental Learning*

Movement was often instigated through the tactile material— participants encouraged students to move, feel, and engage with the material itself. We interpreted movement through explicit mentions of students moving their bodies, arms, hands, or fingers, as well as through implicit mentions of independence and travel. When participants explained the benefits of movement, they referred to the development of their students' fine motor skills, gross motor skills, or spatial concepts. According to participants, fine motor skills referred to the strength and dexterity of the student's hands and fingers, whereas gross motor skills referred to whole or upper body movement. Spatial concepts referred to one's ability to be aware of their own body in relation to the environment. Movement helped students develop motor abilities, build spatial awareness, and comprehend abstract concepts.

To build gross motor ability and spatial concepts, Kamilah set up traffic cones in a classroom (Figure 4E) and guided her student accordingly to learn pedestrian boundaries. Notably, she stated that this practice required, "lots of repetition," for her student to be able to navigate independently. For similar purposes, Tahani employed pattern matching wheels to improve her student's fine motor, gross motor, and spatial awareness (Figure 5C) and Olivia created a tactile chart to convey directions and move her student's fingers up, down, left, and right, to build fine motor and spatial concepts (Figure 4A-C). In the *Interview Study*, Tahani spoke about using "objects that can move from one side to the other [because] you're physically doing it and it stays in your brain longer."

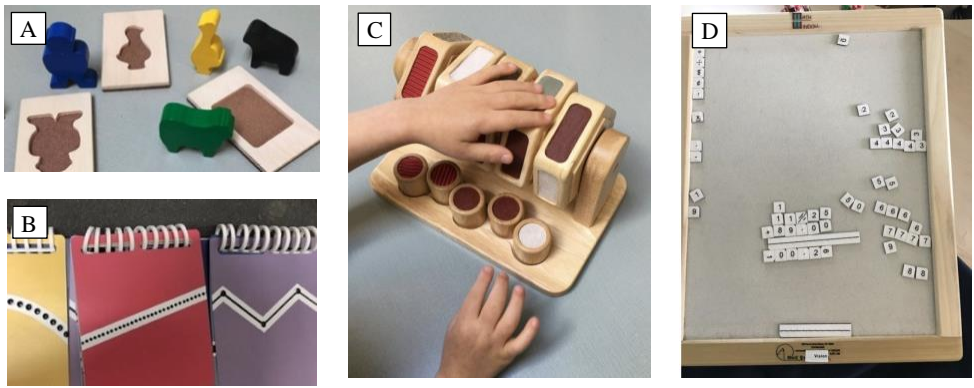

Figure 5: (A) Shape matching, presented by Tahani. (B) Flip-Over Line Paths Concept Books, Olivia noted that students could easily recreate numerous patterns and challenge themselves with more paths. (C) Textured pattern matching wheels. Tahani found that, "the way the puzzle rotated [was] entertaining for kiddos." (D) Math Window- a metal board with numbered magnets to "practice adding numbers with decimals"

Participants incorporated movement through 3D models such as the abacus, or math unit blocks to teach abstract mathematical concepts. Blossom used a board with tactile numbered magnets, called a Math Window, to teach addition (Figure 5D). She explained that "using the math board allows for discussion on spatial math, and manipulating the pieces touches on spatial awareness." She considered letting her student use a braille writer to work through the math problems, but stated that the Math Window would, "help [her] student grasp spatial math concepts better than using a braille writer [and] build spatial awareness skills." Similarly, Alexis used a pegboard with rubber bands to teach geometry (Figure 6A) instead of raised line drawings because it would offer her student "more independence" as she would be able to create the polygons herself. Three other participants used the abacus to teach math, which encouraged students to manipulate the beads and practice their fine motor ability.

*Benefits of Real Objects*

Seven participants presented at least one real object to a student (18 entries, 25.7%). As discussed by participants during the interviews, real objects offered textures that were difficult to replicate through tactile graphics or 3D models, and participants conveyed that real objects (i) served as a "sensory reward" to the student, and (ii) supported the students' understanding of a concept by providing an accurate simulation of the real world.

Connie used six different real objects across her entries, some served as a "[tactile] stimulation [and] reward." In her final interview, she explained that she gave congratulatory rewards to her students in the forms of unique textures and scents. In fact, Connie kept a Sensory Box of objects that she used for tactile stimulation. "[My] students have a very strong desire to touch anything. I am always looking for new things for [them] to touch," she explained, implying that her students were especially excited to feel unique textures. These novel textures were best presented through real objects, such as bubble wrap (Figure 4D) and rubbery drums (Figure 4F).

Real objects also provided students with experience and practice with the real world. Blossom used jars, beads, and a bowl (Figure 3A) to help her student to, "practice pouring from one container into another." Eleanor provided her student with a telephone. She considered using braille labels and stickers to represent the keys and dial pad on the phone but avoided doing so because she wanted her student to "understand the parts of a phone, how to use it, and how to find the numbers tactilely." She explained that "the phone is an example of a school or office phone and the student needs experience dialing and practicing speaking." By minimizing the use of braille labels, Eleanor simulated a real-world scenario. As an improvement, she wanted to connect a working cable to this phone so her student could practice speaking and holding a phone conversation.

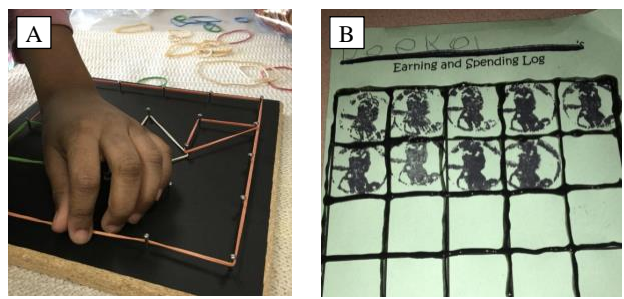

Figure 6: (A) Alexis used a pegboard and rubber bands to "take away the need to practice drawing straight lines and focus on the math concept" of geometry. (B) Worksheet given to all students in the general math classroom. Alexis applied puffy paint to the outlines and worked with the student to complete this sheet.

*Independence and Classroom Inclusion*

As discussed in the interviews, tactile materials fostered independence within students. The diary entries provide examples of materials that encouraged independence, especially in the classroom. Alexis submitted four diary entries that involved tactile material creations with classroom inclusion in mind. "The student prefers to have the same or similar materials as her peers," she explained, so in a particular instance, she added tactile lines onto a printed worksheet that the rest of the class was using (Figure 6B). Four participants made use of materials already supplied for the general classroom like math blocks and 3D shapes. Although the Interview Study spoke to the amount of customization and personalization that tactile materials required, we learn that such unique and exclusive materials carry social implications that may cause students to feel uncomfortable around their peers.

**5 DISCUSSION: REFLECTIONS AND CONSIDERATIONS FOR THE COMMUNITY**

We sought to understand the current practice of tactile materials and gain a deeper understanding of how teachers obtained, created, and presented tactile materials to their students. In our *Interview Study*, teachers reflected on their experiences and explained their challenges with tactile materials. In our *Diary Study*, teachers shared images of the tactile materials they used, and explained their decision-making process for each tactile material. Here, we reflect on findings from *both* studies and situate the perspectives of TVIs with prior research on tactile materials and educational practice.

We draw attention to our participants' reliance on crafty and low-tech materials in their surroundings as opposed to professional tactile material creation tools, which were only used to translate visual material in classrooms for specific academic lessons (i.e., English, Geography). Many teachers steered away from technology to support their creative process; a finding we believe deserves further consideration by the research community. Researchers and designers of assistive technology hold implicit assumptions that incorporating technology into the process of creating and teaching with tactile materials is good, but we question the real needs and practical benefits of introducing new tools. What factors are needed for new and better tools to be adopted? Based on our findings, we discuss design considerations for tools to support tactile material preparation, and to inform the design of assistive technologies to better support TVIs and their students.

### 5.1 Design Tactile Materials for the Expanded Core Curriculum

Prior work on tactile materials has largely focussed on specific subject areas, such as mathematics [21,26,33], computer science [16,25], and the natural sciences [24,54], while our participants covered topics that ranged well beyond these subjects, as outlined in the Expanded Core Curriculum (Section 2.3). TVIs were tasked with creating tactile materials that not only imparted academic knowledge, but also encouraged independent life skills, developed spatial reasoning skills, and instilled values of self-determination and self-confidence within students. "Success in school goes beyond ensuring that students are able to pass their courses and graduate from high school on time. Children and youths with visual impairments deserve the opportunity to have full, rich lives that include good educations, strong social lives, meaningful careers, and the ability to live and travel independently" [47]. Researchers and designers are encouraged to create tactile materials that will help students develop a strong foundation in tactile literacy and spatial reasoning, which will enable personal agency. Materials should employ multiple distinct textures and allow students to explore and self-discover concepts from the tactile material with minimal assistance.

### 5.2 Adapt to a Range of Vision Impairments

All participants created and adapted tactile materials for students with different levels of vision impairments, ranging from no vision to low vision, to "enough vision loss that it impeded [their] ability to learn" (Olivia). Many tactile materials involved braille, whilst most students were unable to read braille and relied on large print (Section 3.2.4). Participants had to adapt tactile materials to suit their students with low vision, and hence sought large, bright, *and* textured materials. Our previous work on tactile maps and 3D objects [49,51,52] addressed this issue by designing models with high-contrast visual feedback. Participants with a range of vision loss were able to learn and use these materials with ease [53]. However, most work has focused on tactile materials for students with no vision. Participants expressed a need for tactile materials that would adapt to different levels of vision, such as being able to easily add tactile cues if their student had no vision or increase the contrast of the colors if they had some residual vision. We encourage future systems to be adaptive to different levels of sight, and appeal to both, tactile and visual senses.

### 5.3 Consider Additional Disabilities

It is crucial to consider the physical and intellectual disabilities that students may have in addition to their visual disability. Prior work on 3D models considered the needs of students with vision impairments [10,21,50–52] but overlooked the maneuvering that would be required by students to properly explore these models (Section 3.2.4). We recommend future developments in 3D models to consider the needs of students with physical disabilities, by offering options for one-hand maneuverability, or by providing alternative input mechanisms for 3D models with digital components.

Students with intellectual disabilities were presented tactile materials as a concept reinforcement and sensory engagement tool. TVIs praised the auditory components of tactile materials, which corroborated prior developments in tactile materials [15,17,52], however, TVIs also sought for new textures and experiences. Future work is needed to understand how students with visual and intellectual disabilities interact with tactile materials, and how tactile materials can better facilitate sensory integration and focus.

### 5.4 Support the Rapid Prototyping and Trial-and-Error Process

Our findings confirmed the need for fast and customizable tools for tactile materials [36,53]—teachers often had limited time to create tactile materials and they preferred materials they could quickly shape (i.e., swell form) and easily customize to reuse for future lessons (Section 4.2.5). Almost half the tactile materials reported were created by participants, demonstrating a tendency and need for rapid prototyping. Expanding on prior work, we highlight the TVIs' need for rapid iteration of tactile materials. Tahani, one of the only participants who had access to 3D printing technology, marveled at the speed and durability of 3D printed materials; but the "trial and error [to] individualize materials" for her student, caused 3D printing to become a costly and inefficient endeavour. Many teachers spoke to the notion of an "afterthought" in the tactile material creation process. Based on the student's response and grasp of the intended concept, they would alter the material; they used materials like clay or Lego blocks because they were easily mouldable and could also afford different textures. Thus, technologists must continue working on rapid prototyping tools for tactile materials but ensure cost-efficient iteration and customization capabilities.

### 5.5 Craft Multisensory Experiences through Tactile, Auditory, and Visual Stimuli

Teachers presented tactile materials while providing oral explanations, auditory effects, visual stimuli (for students with low vision), and olfactory stimuli (Section 3.2.2). These multisensory experiences engaged different senses of the student and helped students to better engage with the material. Although tactile stimuli might be considered trivial in the context of a tactile material, teachers wished for greater tactile distinction and contrast in their tactile materials (Section 3.2.4, 4.2.4). In some cases, recognizing different textures (Figure 5C) and overcoming the student's tactile sensitivity was the goal of the lesson.

Previous work on tactile materials recognized the importance of audio and have integrated audio annotations in tactile graphics [4], 3D models [52] and tactile maps [39,72]. However, in many cases, audio was more than annotations or explanations, audio effects could help students build a spatial compass and develop their spatial reasoning skills. Audio effects were also enjoyable, fun, and could deliver feedback about a student's progress.

Tactile stimuli need to be easily distinguishable with distinct textures, auditory stimuli need to be engaging and consistent, and visual stimuli must have high contrast. Future work should combine tactile, auditory, and other stimuli depending on the student's preferences to ensure a conducive learning experience.

### 5.6 Use 3D Models to Build Independence and Add Saliency to Tactile Graphics

Tactile graphics have been researched extensively in prior literature [2,48,69–71], whereas there has been little consideration, except for Holloway et al.'s [20], and our previous work [50–53], on the critical role of 3D models and real objects as tactile materials. By themselves, 3D models were ideal teaching material for students in elementary grades as they closely resembled the real world. According to teachers, many students explored 3D models to develop a conceptual understanding of the topic and build independent learning skills (see Section 3.2.3).

When used with tactile graphics, 3D models highlighted salient elements of the graphic by adding a layer of depth, allowing students to process information much more easily, and resolving issues surround tactile graphic overload [2,45,71]. Tactile graphics were an abstract hump that students were not always ready to conquer (Section 4.2.2), and many students were not prepared to understand tactile graphics until their later years. Through our *Diary Study*, we found that many tactile materials were a combination of tactile graphics *and* 3D models. Participants used 3D models alongside science diagrams (Figure 2D) and tactile book illustrations (Figure 2B), referred to as story box objects [66] which supported the student's understanding of the narrative. Previous technical innovations advanced tactile graphics [5,16,32,42] and 3D models [50–52] as unique entities. However, we discovered that in an ideal scenario, teachers preferred to present tactile graphics *with* 3D models.

Despite the necessity of 3D models, all TVIs found 3D models difficult to create or obtain, and most did not have access to a 3D printer. In addition, 3D models were obtrusive and difficult to carry around, which negated the independence they could afford. As a result, teachers frequently resorted to a tactile graphic as it resembled the materials of sighted students (Figure 6B) and were relatively easier to create and modify. Future technologies must support the use of 3D models and create tools that would allow economical and iterative modeling of 3D tactile materials. Multipart tactile graphics, such as diagrams and scenery should employ 3D models to emphasize certain parts. However, 3D models for general education subjects must be discreet and avoid drawing unwanted attention.

**5.7 Promote Movement through Tactile Materials**

Participants gravitated towards materials that encouraged their students' upper body movement or intricate movement of their fingers. They sought such materials to develop fine motor and gross motor skills within students, to build upon the student's spatial awareness, and to convey abstract concepts, such as math.

Findings from our diary study expand upon past literature in Embodied Learning (Section 2.3). We illustrate how TVIs intentionally presented movable tactile materials to develop their students' motor abilities and spatial awareness, such as the textured pattern matching wheels (Figure 5C) and story box objects (Figure 2B). Materials such as the abacus, Math Window (Figure 5D), and the geometry pegboard (Figure 6A) reinforced the importance of movement and object manipulation in teaching math. Most participants in our study used object manipulatives for math (see Section 4.2.5).

Future assistive systems should consider ways to promote movement and allow students to develop their motor abilities and spatial concepts. In addition, future tactile materials for math instruction should involve interactive and physical elements for object manipulation.

**5.8 Support Student-Teacher Tactile Material Co-Creation**

Three participants involved students in their tactile material creation process. Although this was a minority of participants, we believe that this practice of co-creation held promise for future work in tactile material creation. Prior work in this space either removed the creative experience altogether by automating tactile material creation [32,42,61], or only considered teachers as the users of their system [49,51,52]. This follows the instruction pattern of traditional educational settings, where teachers assumed the role of experts who relayed a one-way knowledge stream to students, while students were tasked with absorbing this knowledge in a passive manner.

Co-creation repositions these roles and shifts this dynamic to an active learning model where students are regarded as *partners* in the teaching and learning experience. We expand upon the pedagogical approach of Students as Partners (Section 2.3) to include students with vision impairments. While prior work on participatory design has involved users with vision impairments [59,63,68], this practice is broadly overlooked in education settings except for Pires et al.'s [43] recent work on creating math tools for students, *with* students and TVIs.

Kamilah, Elizabeth, and Tahani engaged their students in the creation process (Section 3.2.1). Elizabeth's student initially disregarded her previously created tactile maps but felt confident to navigate the school independently after co-creating a map with Elizabeth. Likewise, Tahani's student felt excited and proud to be able to create his own tactile material (Figure 3C). The experiences of our participants lead us to understand that co-creation, involving SaP, nurtured values of self-determination and independence within students. We suggest that technologies should allow a means for student and teacher collaboration, not just for student participation, but for active student involvement and decision-making, so that they can help to guide the creation process of the materials that resonate with their learning preferences and abilities.

### 5.9 Facilitate an Online Social Network

Previous research in general education teacher collaboration suggests that social networks were helpful in combating frustration and isolation, while also inspiring novel instruction methods [35]. In particular, social media platforms were incredible repositories of educational resources [44]. Allowing teachers to engage as a consumer, a networker, or as a content creator, social media platforms facilitated collaboration and helped contribute new ideas to advance educational practice [44]. Per our findings, most TVIs collaborated with the staff at their students' schools, but rarely went online to network or gather ideas, despite its benefits. We noted an incident where one TVI purchased an APH tactile clock and found difficulties in using it, while another TVI knew that the APH clock would not suit her student's needs and created her own tactile clock (Figure 1C). A sharing of ideas and experiences could have helped the first TVI to make an informed decision about purchasing the APH tactile clock.

To our knowledge, there has been no considerable work to understand how TVIs share ideas, nor has there been developments in technologies to ease this process. The APH Tactile Graphic Image Library provides well-designed templates for tactile graphics, and a Pinterest search can offer multiple tactile material ideas. Thingiverse.com has also emerged as an informal social media platform where users regularly shared their 3D printed and modified assistive devices [9].

Though tactile material creation can be very student dependent, we see a large opportunity in this space to facilitate TVI social networks, and we encourage researchers to consider how TVIs may share their tactile material creations and insights with others.

## 6 LIMITATIONS

The goal of this study was to understand how TVIs create and use tactile materials in practice. While the 21 interviews were incredibly thorough, we understand that participants may have recalled memorable insights rather than information pertaining to their regular use of tactile materials. We mediated this challenge by conducting a 3-week diary study and asking participants to report information about every tactile material they created. However, we have limited knowledge about the absolute number of tactile materials produced and used by TVIs in the 3-week time frame. We accompanied diary entries with follow-up interviews to clarify any missed or misrepresented information in the diary study. In addition, we situated our results and design considerations within existing literature on tactile materials.

The majority of our participants were from Washington and New York, and we understand that states have different educational districts and policies which may affect the way TVIs work with students. In addition, we focus on the benefits and critiques of tactile material instruction from the perspective of TVIs, however, we acknowledge that we would need to gather input from students with vision impairments to understand the true learning experiences that tactile materials provide. Aldrich and Sheppard [2] noted that older students were particularly likely to give useful design feedback. With the perspectives of both teachers and students, we would have a broader understanding of the use of tactile materials in the classroom. Future studies can explore a longer timeframe and consider a diverse range of participants to gather findings that are widely applicable and generalizable.

## 7 CONCLUSION

Through semi-structured interviews and diary studies, we shed light on social and contextual aspects of tactile material use. We then used these insights to offer design considerations for future work on tactile materials. These insights also highlighted the mismatch between current technical advancements in tactile material design, and the current practice of tactile materials by TVIs and students. Our findings contribute to the following key areas:

- Role of tactile materials within a school setting
- Current acquisition and creation practices of tactile materials
- Contrasts between different types of tactile materials (tactile graphics, 3D models, real objects)
- Social and cognitive benefits of tactile materials (e.g., independence, classroom inclusion)

Although recent advances in tactile materials have focussed on conveying academic concepts, our findings dictate that teachers also used tactile materials for conveying non-academic concepts. During one-on-one sessions, students were presented tactile materials to develop their tactile literacy, motor abilities, and spatial reasoning abilities, confirming the existing literature on the ECC. We discovered that these skills were best conveyed through 3D models and real objects, which all TVIs preferred, but found difficult to obtain or create.

Fundamentally, the implicit goal behind each tactile material was to foster independence. By enabling multisensory experiences, movement, student collaboration, TVIs nurtured values of self-advocacy and determination within students, and remarkably, preferred low-tech, reliable solutions to ease their rapid tactile material creation process. Given the growing interest in assistive technologies for tactile material creation, our research provides a fundamental understanding of current practices, highlighting considerations and opportunities for future research.

## ACKNOWLEDGMENTS

This research was supported in part by Verizon Media Group and the National Science Foundation (NSF) under grant No. IIS-1746123. We thank Yilin Xu, Sindhu Banavara Ravindra, and Bhagyasri Canumalla. We also thank all participants for their time and helpful insights.